\documentclass[twocolumn]{aa}  
\usepackage{graphicx,natbib}
\usepackage{txfonts}
\newcommand{\Msun}{\ensuremath{M_{\odot}}}

\newcommand{\re}{\ensuremath{r_{\rm e}}}
\newcommand{\Hb}{\ensuremath{{\rm H}\beta}}

\newcommand{\Mgb}{\ensuremath{{\rm Mg}\, b}}

\newcommand{\aFe}{\ensuremath{\alpha/{\rm Fe}}}

\newcommand{\ZH}{\ensuremath{Z/{\rm H}}}

\begin{document}
   \title{A counter-rotating core in the dwarf elliptical galaxy VCC 510}

   \subtitle{}

   \author{D. Thomas
          \inst{1,2}
          \and
          F. Brimioulle\inst{2,3}\and R. Bender\inst{2,3}\and
	  U. Hopp\inst{2,3} \and L. Greggio\inst{4}\and
	  C. Maraston\inst{1,2} \and R.P. Saglia\inst{2} }

   \offprints{D. Thomas}

   \institute{University of Oxford, Denys Wilkinson Building, Keble
              Road, Oxford, OX1 3RH, UK\\
              \email{dthomas@astro.ox.ac.uk}
	      \and
             Max-Planck-Institut f\"ur extraterrestrische Physik,
             Giessenbachstra\ss e, 85748 Garching, Germany
	     \and
	     Universit\"ats-Sternwarte M\"unchen, Scheinerstr.\ 1,
             81679 M\"unchen, Germany
	     \and
	     INAF, Osservatorio Astronomico di Padova, vicolo
	     dell'Osservatorio 5, 35122 Padova, Italy
             }

   \date{Received October 27, 2005; accepted November 13, 2005}

 
  \abstract 
  {}
%
  {We present optical long-slit spectra of the Virgo dwarf elliptical
  galaxy VCC 510 at high spectral ($\sigma\sim 30\;$km/s) and spatial
  resolution. The principal aim is to unravel its kinematical and
  stellar population properties.}
  {Heliocentric velocities and velocity dispersions as functions of
  galaxy radius are derived by deconvolving line-of-sight velocity
  distributions.  The luminosity-weighted stellar population
  parameters age and element abundances are obtained by comparison of
  Lick absorption-line indices with stellar population models.}  
%
  {A maximum rotation $v_{\rm rot}=8\pm 2.5$ km/s inside half the
  effective radius ($\re\approx 20\arcsec$) and a mean, radially flat
  velocity dispersion $\sigma=44\pm 5$ km/s are measured. The core
  extending over the inner 2\arcsec\ ($\sim 140\;$pc) is found to
  rotate in the opposite sense with $v_{\rm rot}^{\rm core}\approx
  -1/2\ v_{\rm rot}$. VCC~510 ($M_B\sim -15.7$) is therefore by far
  the faintest and smallest galaxy with a counter-rotating core
  known. From the main body rotation and the velocity dispersion
  profile we deduce that VCC 510 is anisotropic and clearly not
  entirely supported by rotation.  We derive an old
  luminosity-weighted age ($10\pm 3\;$Gyr) and sub-solar metallicity
  ($[\ZH]=-0.4\pm 0.1$) inside the effective radius. There is
  tentative evidence that the counter-rotating core might be younger
  and less \aFe\ enhanced. From the stellar population parameters we
  obtain a total stellar mass-to-light ratio of $\sim
  3.6\;(M/L_B)_{\odot}$ which is significantly lower than a rough
  dynamical estimate obtained from the kinematics through the virial
  theorem ($\sim 15$). This discrepancy hints toward the possible
  presence of dark matter in the centre of VCC 510.}
%
%
  {We discuss the origin of the counter-rotating core and exclude
  fly-by encounters as a viable possibility. Gas accretion or galaxy
  merging provide more likely explanations. VCC 510 is therefore the
  direct observational evidence that such processes do occur in
  cluster satellite galaxies on dwarf galaxy scales.}

   \keywords{Galaxies: dwarf --
             galaxies: kinematics and dynamics --
             galaxies: formation --
	     galaxies: interactions --
	     galaxies: stellar content               }

   \maketitle
%

\section{Introduction}
Counter-rotating cores are found in a considerable fraction of giant
early-type galaxies
\citep{Be88,FI88,JS88,BS92,Beretal94,Mehetal98,WEC02,Emsetal04}. Objects
with this kinematic peculiarity constitute a subclass of an even more
common group of galaxies hosting so-called kinematically decoupled
components (KDCs). First discovered by \citet{EEC82}, the latter are
cores that rotate at different speed from the main body of the
galaxy. Their formation, and in particular the formation of
counter-rotating cores, is best understood in a merging scenario,
hence their presence is considered to be very direct evidence that
merging plays a r\^ole in the formation of giant early-type galaxies
\citep{Kormendy84,BQ90}.

\begin{figure*}
\centering
\includegraphics[width=0.78\textwidth]{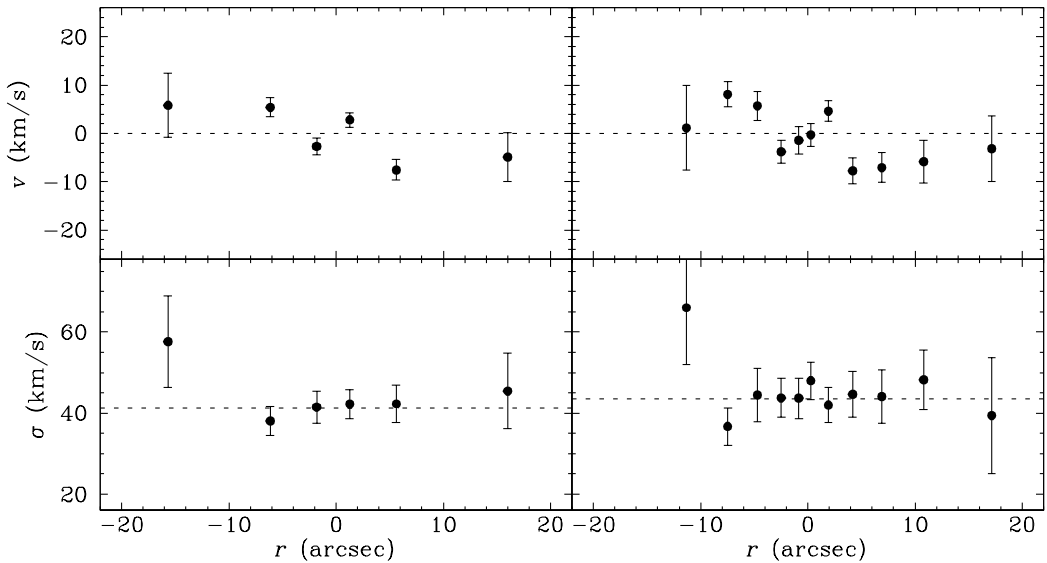}
\caption{Heliocentric velocity $v$ (mean $v$ subtracted) and velocity
dispersion as functions of radius. Coarse and fine binnings are shown
by the left-hand and right-hand panels, respectively. 1\arcsec\
corresponds roughly to 70$\;$pc assuming $14.6\;$Mpc for the distance
of the Virgo cluster \citep{Freetal01}.}
           \label{fig:rotcurve}%
\end{figure*}

Since its discovery in the late 80's, this interesting kinematical
feature has been restricted to massive, luminous galaxies. For our
understanding of galaxy formation, it is certainly of great interest
to know whether this limitation is of pure observational nature and
whether merger remnants with these kinematical fingerprints exist also
on smaller galaxy scales. A first hint that this might be the case
comes from recent observations by \citet{GGvdM05}, who found a
counter-rotating core in the low-luminosity elliptical NGC~770, being
the faintest galaxy ($M_B\sim -18$) with this kinematical feature
known at that time \citep[see also][]{Pruetal05}.

First evidence that the presence of kinematical peculiarities in
galaxy centres extends even further down to dwarf galaxy scales is
provided by \citet{DeRijetal04}. The authors have detected kinematic
decoupling in the centres of two dwarf elliptical galaxies, where in
both cases a pronounced bump in the rotation velocity profile at a
radius of 1\arcsec\ ($\sim 200\;$pc) is found. The cores of these
objects rotate in the same sense and with similar velocities as the
main body. \citet{DeRijetal04} argue that this feature is best
understood in a harassment scenario and is most likely caused by
fly-by interaction with the nearby massive galaxy.

In this paper we report on the detection of a counter-rotating core in
a dwarf elliptical galaxy. We present the kinematics of VCC~510, a
nucleated dwarf elliptical of the Virgo cluster
\citep{BST85,BiCa91,BiCa93} with moderate flattening
($\epsilon=0.18$), a total blue luminosity of $M_B\sim -15.7$, and an
effective radius of $r_{\rm e}=20\arcsec\sim 1.4\;$kpc. We assume
$14.6\pm 0.04\;$Mpc for the distance to the Virgo cluster
\citep{Freetal01}. Being more than two magnitudes fainter than
NGC~770, VCC~510 is by far the smallest galaxy known with a
counter-rotating core.


\section{Observations and data analysis}
The observations were performed at the 3.5m telescope (TWIN
spectrograph) of the Calar Alto Observatory (Spain) in April 2003. A
long-slit ($1.8\arcsec$ width) was placed at the centre of VCC~510
aligned along the major axis, yielding a wavelength coverage $4650\la
\lambda/{\rm \AA}\la 5600$ at relatively high instrumental resolution
($\sigma_{\rm inst}\sim 30\;$km/s). A total exposure time of 5.5 hours
yielded a signal-to-noise ratio (\AA$^{-1}$) of 30 for fine and 40 for
coarse radial binning (see Fig.~\ref{fig:rotcurve}) inside \re, with a
seeing of about 2\arcsec. We also observed 20 standard stars serving
as templates for both the kinematic analysis and the calibration of
Lick absorption-line measurements (see below). VCC~510 is one of the
35 dwarf early-type galaxies mostly from the Virgo cluster we have
observed over the past 10 years. The complete sample will be presented
in a future paper, full details about data reduction will be given
there. In the following we briefly summarise the essentials.

The standard CCD data reduction (bias \& dark subtraction, flat
fielding, wavelength calibration, and sky subtraction) was carried out
under the image processing package MIDAS provided by
ESO. Line-of-sight-velocity-distributions were determined by using the
Fourier-Correlation-Quotient (FCQ) method from \citet{Bender90a}, with
which we derived heliocentric velocities and velocity dispersions
\citep{BSG94}. We measured the Lick absorption-line indices \Hb, \Mgb,
Fe5270, and Fe5335 \citep{Woretal94}. For this purpose, the spectra
were degraded to Lick resolution, and the intrinsic velocity
broadening, even though almost negligible in case of dwarf galaxies,
was taken into account. The measurements are calibrated onto the Lick
system by means of the 20 Lick standard stars observed. For both
kinematics and line indices, statistical errors were derived from
Monte Carlo simulations.

\section{Results}
\begin{figure*}
\centering
\includegraphics[width=0.4\textwidth]{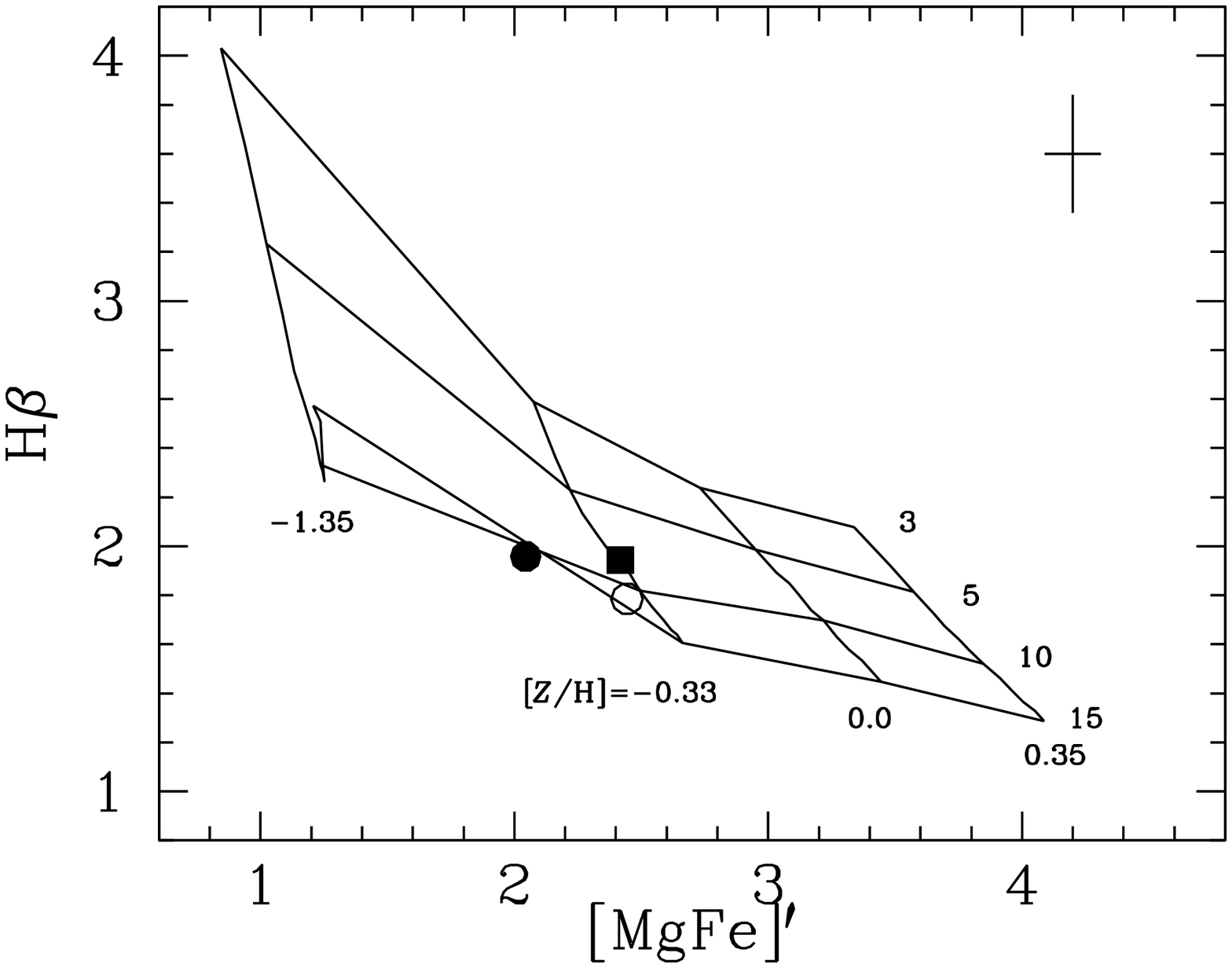}
\includegraphics[width=0.4\textwidth]{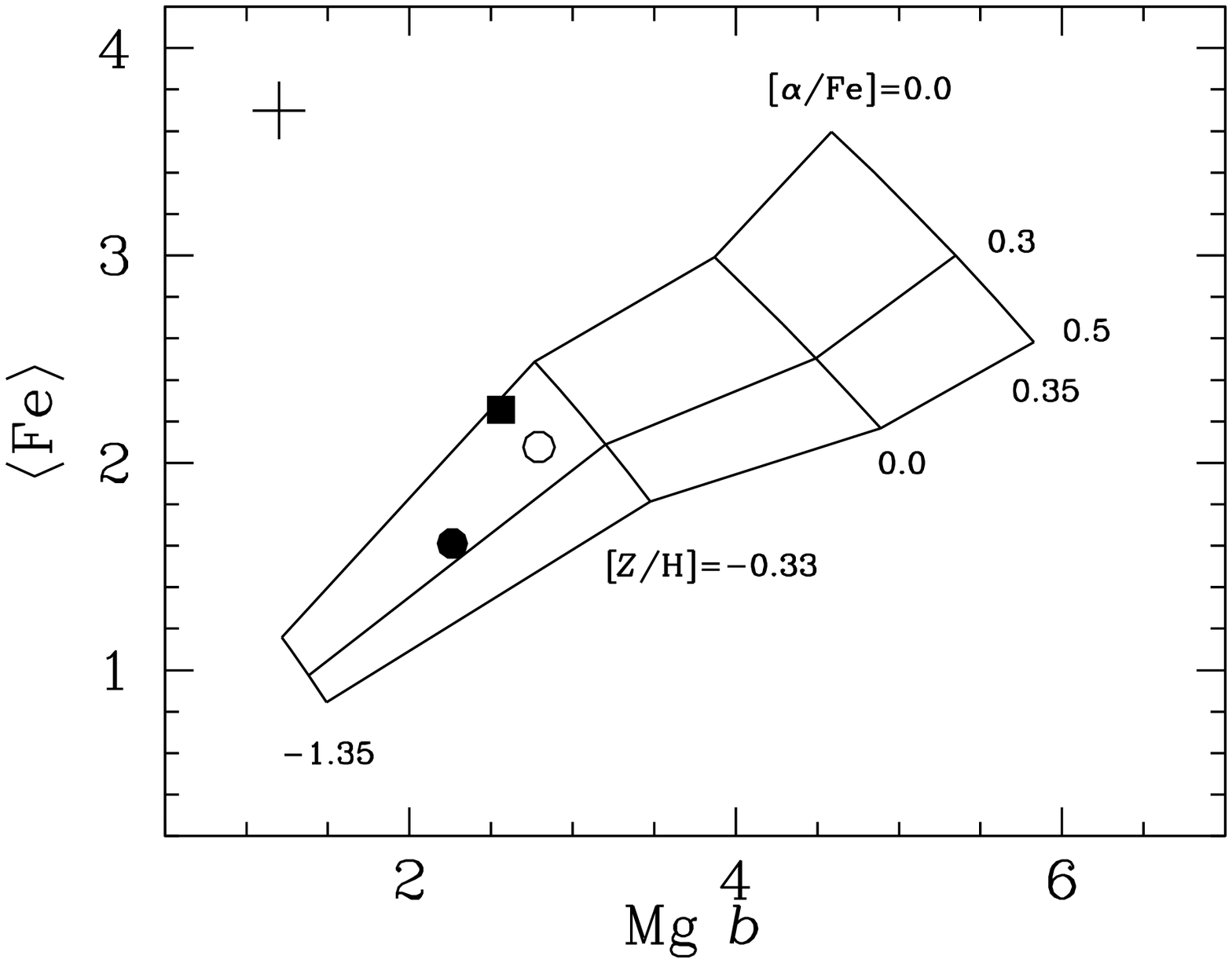}
\caption{Lick absorption-line indices for three radial bins: very
centre (filled circle), at $r=2\arcsec$ (filled square, radius of the
counter-rotating core), and at $r=5\arcsec$ (open circle). Stellar
population models \citep{TMB03a} for various ages, metallicities, and
\aFe\ ratios (see labels) are over-plotted. In the left-hand panel
\aFe\ is fixed to solar, in the right-hand panel age is fixed to
10$\;$Gyr.  For the definition of the indices see \citet{TMB03a}.}
\label{fig:indices}
\end{figure*}
\begin{table*}
\begin{minipage}[t]{\columnwidth}
\caption{Kinematical parameters of VCC 510}
\label{tab:parameters}
\centering
\renewcommand{\footnoterule}{}
\begin{tabular}{ccccccccccc}
\hline\hline
$M_B$ & $r_{\rm e}$ & $\epsilon$ & $v_{\rm hel}$ & $v_{\rm rot}$ &$v_{\rm rot}^{\rm core}$ & $\sigma$ & $v/\sigma$  & $M_{\rm dyn}$ & $(M/L_B)_{\rm dyn}$\\
(mag) & (arcsec) & ~ & (km/s) & (km/s) & (km/s) & (km/s) & ~ & ($10^9\Msun$) & $(M/L_B)_{\odot}$\\
\hline
$-15.7$ & 20 & 0.18 & $858\pm 1$ & $8.0\pm2.5$ & $\sim -1/2\ v_{\rm rot}$ & $44\pm 5$ & $0.18\pm 0.06$ & $4.5\pm 0.5$ & $15\pm 2$\\
\hline
\end{tabular}
\end{minipage}
\end{table*}
\subsection{Kinematics}
The rotation curve and the radial velocity dispersion profile of
VCC~510 are shown in Fig.~\ref{fig:rotcurve} for coarse (left-hand
panel) and fine (right-hand panel) radial binning. In both cases, but
in particular for the fine binning, the counter-rotating core is
clearly visible. The core region inside 2\arcsec\ ($\sim 140\;$pc) is
rotating with about $4\;$km/s in a sense opposite to the main body
rotation. We are confident that this feature is real, as it is robust
against variations of the radial binning as shown in
Fig.~\ref{fig:rotcurve}. Note also that the rotation velocity of the
core should be regarded as a lower limit, because the moderate seeing
of the present observations contributes to a dilution of the
signal. The maximum rotation velocity of the galaxy (at $\sim
6\arcsec$) is $v_{\rm rot}=8\pm 2.5\;$km/s.

The velocity dispersion of VCC~510 is $\sigma=44\pm5\;$km/s and
reveals no radial dependence inside $r_{\rm e}$, which is a typical
characteristic for dwarf elliptical galaxies
\citep{BN90,BPN91,GGvdM03}. The resulting relatively low
$v/\sigma=0.18\pm 0.06$ at the ellipticity $\epsilon=0.18$ shows that
VCC~510 is supported by anisotropic velocity dispersion
\citep{Binney05,BN05}.  A significant increase of $v_{\rm rot}$ beyond
$r_{\rm e}$, as found by \citet{SP02} for NGC~205, cannot be excluded,
but seems unlikely. Fig.~\ref{fig:rotcurve} indicates even a slight
decrease of the rotation velocity at the outermost point.

Finally, following the recipe of \citet{BT87}, we obtain the virial
mass $M_{\rm vir}=4.5\pm 0.5 \times 10^9\;$\Msun, which can be
considered a rough estimate of the total dynamical mass. This leads to
a relatively high dynamical $B$-band mass-to-light ratio of $15\pm 2$
$(M/L_B)_{\odot}$. The results are summarised in
Table~\ref{tab:parameters}.

\subsection{Stellar populations}
\label{sec:ssp}
In Fig.~\ref{fig:indices} we show the Lick absorption-line indices
measured for three radial bins and confront them with stellar
population models of various ages, metallicities, and \aFe\ ratios as
indicated by the labels \citep{TMB03a}. Using the method
explained in detail in \citet{Thoetal05}, we derive the following
global (within 5\arcsec) luminosity-weighted stellar population
parameters: age $t\approx 10.3\pm 2.8\;$Gyr, metallicity
$[\ZH]=-0.4\pm 0.13$, and abundance ratio $[\aFe]=0.07\pm 0.08\;$dex.
Most importantly in the context of this letter, the counter-rotating
core does not show any clear peculiarity in the stellar population
properties within the measurement errors, in particular no obvious
sign of recent star formation. This seems indeed to be a quite common
characteristic for counter-rotating cores in elliptical galaxies
\citep{SB95,Mehetal98,Davetal01,Moretal04}.

However, a closer look at the figure suggests that the stellar
populations at the radius of the counter-rotating core (square) might
be somewhat younger by about $2\;$Gyr, and less alpha/Fe enhanced by
$\sim 0.2\;$dex. But, given the measurement errors, this result can be
considered indicative at best. Assuming the core contributes about
$10\,$--$\,30$ per cent in mass (enough to produce counter-rotating
cores, see Surma \& Bender 1995), this slight rejuvenation in the core
region would imply the presence of a young component with age $t\sim
4\,$--$\,6\;$Gyr, corresponding to formation redshifts of $z\sim
0.5\,$--$\,0.8$.

Based on the stellar population models of \citet{Ma05}, the global
mean age and metallicity derived above translate into the stellar
mass-to-light ratio $(M/L_B)_*=3.6\pm 0.7\;$ $(M/L_B)_{\odot}$
assuming a Kroupa initial mass function (Salpeter increases the M/L by
a factor $\sim 1.6$).  This value is significantly smaller than the
dynamical estimate presented above. Unless the simple recipe used here
over-estimates the mass by more than a factor 3, this indicates that
dark matter might be present in the centre of VCC~510. Sophisticated
modelling based on more observational data is certainly needed for a
detailed assessment.

\section{Discussion and Conclusion}
Given the evidence presented here, there is no doubt that VCC~510
contains a counter-rotating core. The challenge is now to understand
(and model) the origin of this kinematic peculiarity. It is widely
accepted that galaxy mergers are the key process to form kinematically
decoupled cores in massive galaxies (see Introduction and references
therein). Is this scenario valid also for dwarfs?

\citet{DeRijetal04} argue that mergers between small galaxies are
generally unlikely. In a dense environment velocities are too high,
and in looser environments densities are too low. They suggest an
alternative explanation, according to which the internal kinematics of
a dwarf galaxy are modified by fly-by encounters with a massive
companion. Developing a simple analytical description, they estimate
that this harassment process can produce a kinematic decoupling in the
centres of dwarf ellipticals, as long as impact parameters of the
interaction are not larger than $\sim 20\;$kpc.

While it might indeed be plausible that fly-by interaction affects
internal kinematics leading to perturbations of the inner core as
found in \citet{DeRijetal04}, it seems highly unlikely that the sense
of rotation can be completely reversed by this mechanism. Adapting
Eqn.~7 of \citet{DeRijetal04}, we find no sensible combination of
parameters (velocity dispersion in Virgo $\sim 600\;$km/s; companion
mass around $10^{11}\;$\Msun; impact parameter $\ga 20\;$kpc) that
would affect the rotation velocity significantly enough to explain the
present observations. \citet{GGvdM05} come to the same conclusion for
the low-mass elliptical NGC~770.

A galaxy merger or gas accretion are the alternatives.  According to
the commonly accepted scenario, a dissipationless, unequal-mass merger
can produce a kinematically decoupled core \citep{Kormendy84,BQ90},
under the premise that the nucleus of the accreted galaxy is
denser. This is plausible for giants, as surface brightness indeed
increases with decreasing galaxy mass, but is a bit problematic for
dwarfs, where the opposite relation is observed
\citep[e.g.,][]{Kormendy85}. More compelling might be a dissipational
merger or gas accretion accompanied by star formation. The
counter-rotating core of NGC~4365, for instance, has formed most
probably through such a process, as it is found to have disc-like
structure, low kinematic temperatures and a significantly flattened
density distribution \citep{SB95}.

The latter scenario gets support from the indication that the
counter-rotating core might have experienced recent star formation as
discussed in Section~\ref{sec:ssp}. In this case, the core should
exhibit a disc-like structure.  Clearly, photometry and spectroscopy
at much higher spatial resolution than presented here are needed to
verify the presence of such a cold component in the centre.

Mergers between dwarfs, on the other hand, are quite unlikely
in clusters given their low cross-section and the high relative
velocities.  Based on N-body simulations by \citet{MH97} and adopting
the dwarf galaxy surface density of Virgo from \citet{Phietal98}, we
estimate that only 1 out of $\sim 400$ dwarf ellipticals in the Virgo
cluster should have experienced a merger over a Hubble time.

VCC~510 is the direct evidence that gas accretion or mergers between
satellites do occur on dwarf galaxy scales. They might be important
for galaxy formation modelling \citep{Menetal02}, and it needs to be
assessed in future how common this phenomenon really is.

\begin{acknowledgements}
We are grateful to the Calar Alto staff for the observations, to J.\
Binney, R.\ Davies, and S.\ Khochfar for interesting discussions, and
to E.\ Emsellem for the very quick and careful report. DT acknowledges
financial support by grant BMBF-LPD 9901/8-111 of the Deutsche
Akademie der Naturforscher Leopoldina.
\end{acknowledgements}

\end{document}